\documentclass[sigconf,screen]{acmart}
\usepackage{indentfirst}
\usepackage{multirow}
\usepackage{tikz}
\usepackage{pgfplots}
\usepgfplotslibrary{groupplots}
\usepackage{pgf-pie}
\usepackage{tcolorbox}
\AtBeginDocument{%
  \providecommand\BibTeX{{%
    \normalfont B\kern-0.5em{\scshape i\kern-0.25em b}\kern-0.8em\TeX}}}

\makeatletter
\newcommand*{\rom}[1]{\expandafter\@slowromancap\romannumeral #1@}
\makeatother

\setcopyright{rightsretained}
\acmPrice{}
\acmDOI{10.1145/3540250.3549099}
\acmYear{2022}
\copyrightyear{2022}
\acmSubmissionID{fse22main-p159-p}
\acmISBN{978-1-4503-9413-0/22/11}
\acmConference[ESEC/FSE '22]{Proceedings of the 30th ACM Joint European Software Engineering Conference and Symposium on the Foundations of Software Engineering}{November 14--18, 2022}{Singapore, Singapore}
\acmBooktitle{Proceedings of the 30th ACM Joint European Software Engineering Conference and Symposium on the Foundations of Software Engineering (ESEC/FSE '22), November 14--18, 2022, Singapore, Singapore}

\begin{document}

\title{AUGER: Automatically Generating Review Comments with Pre-training Models}

\author{Lingwei Li}
\affiliation{%
  \institution{Institute of Software, CAS, Univ. of Chinese Academy of Sciences}
  \city{Beijing}
  \country{China}}
\email{lilingwei20@mails.ucas.ac.cn}

\author{Li Yang}
\authornote{Corresponding author}
\affiliation{%
  \institution{Institute of Software, CAS}
  \city{Beijing}
  \country{China}}
\email{yangli2017@iscas.ac.cn}

\author{Huaxi Jiang}
\affiliation{%
  \institution{Institute of Software, CAS, Univ. of Chinese Academy of Sciences}
  \city{Beijing}
  \country{China}}
\email{jianghuaxi19@mails.ucas.ac.cn}

\author{Jun Yan}
\affiliation{%
  \institution{State Key Laboratory of Computer Science, Institute of Software, CAS, Univ. of Chinese Academy of Sciences}
  \city{Beijing}
  \country{China}}
\email{yanjun@ios.ac.cn}

\author{Tiejian Luo}
\affiliation{%
  \institution{Univ. of Chinese Academy of Sciences}
  \city{Beijing}
  \country{China}}
\email{tjluo@ucas.ac.cn}

\author{Zihan Hua}
\affiliation{%
  \institution{Wuhan University, Univ. of Chinese Academy of Sciences}
  \city{Wuhan}
  \country{China}}
\email{2018302110434@whu.edu.cn}

\author{Geng Liang}
\affiliation{%
  \institution{Institute of Software, CAS}
  \city{Beijing}
  \country{China}}
\email{lianggeng@iscas.ac.cn}

\author{Chun Zuo}
\affiliation{%
  \institution{Sinosoft Company Limited}
  \city{Beijing}
  \country{China}}
\email{zuochun@sinosoft.com.cn}



\begin{abstract}
Code review is one of the best practices as a powerful safeguard for software quality. In practice, senior or highly skilled reviewers inspect source code and provide constructive comments, considering what authors may ignore, for example, some special cases. The collaborative validation between contributors results in code being highly qualified and less chance of bugs. However, since personal knowledge is limited and varies, the efficiency and effectiveness of code review practice are worthy of further improvement. In fact, it still takes a colossal and time-consuming effort to deliver useful review comments.

This paper explores a synergy of multiple practical review comments to enhance code review and proposes \textbf{AUGER} (\textbf{AU}tomatically \textbf{GE}nerating \textbf{R}eview comments): a review comments generator with pre-training models. We first collect empirical review data from 11 notable Java projects and construct a dataset of 10,882 code changes. By leveraging Text-to-Text Transfer Transformer (T5) models, the framework synthesizes valuable knowledge in the training stage and effectively outperforms baselines by \textbf{37.38\%} in ROUGE-L. \textbf{29\%} of our automatic review comments are considered useful according to prior studies. The inference generates just in 20 seconds and is also open to training further. Moreover, the performance also gets improved when thoroughly analyzed in case study.
\end{abstract}

\ccsdesc[500]{Software and its engineering~Software creation and management}
\ccsdesc[500]{Computing methodologies~Machine learning}

\keywords{Review Comments, Code Review, Text Generation, Machine Learning}

\maketitle
\section{Introduction}
Modern code review is considered effective in reducing 
\begin{figure}[h]
  \centering
  \includegraphics[width=\linewidth]{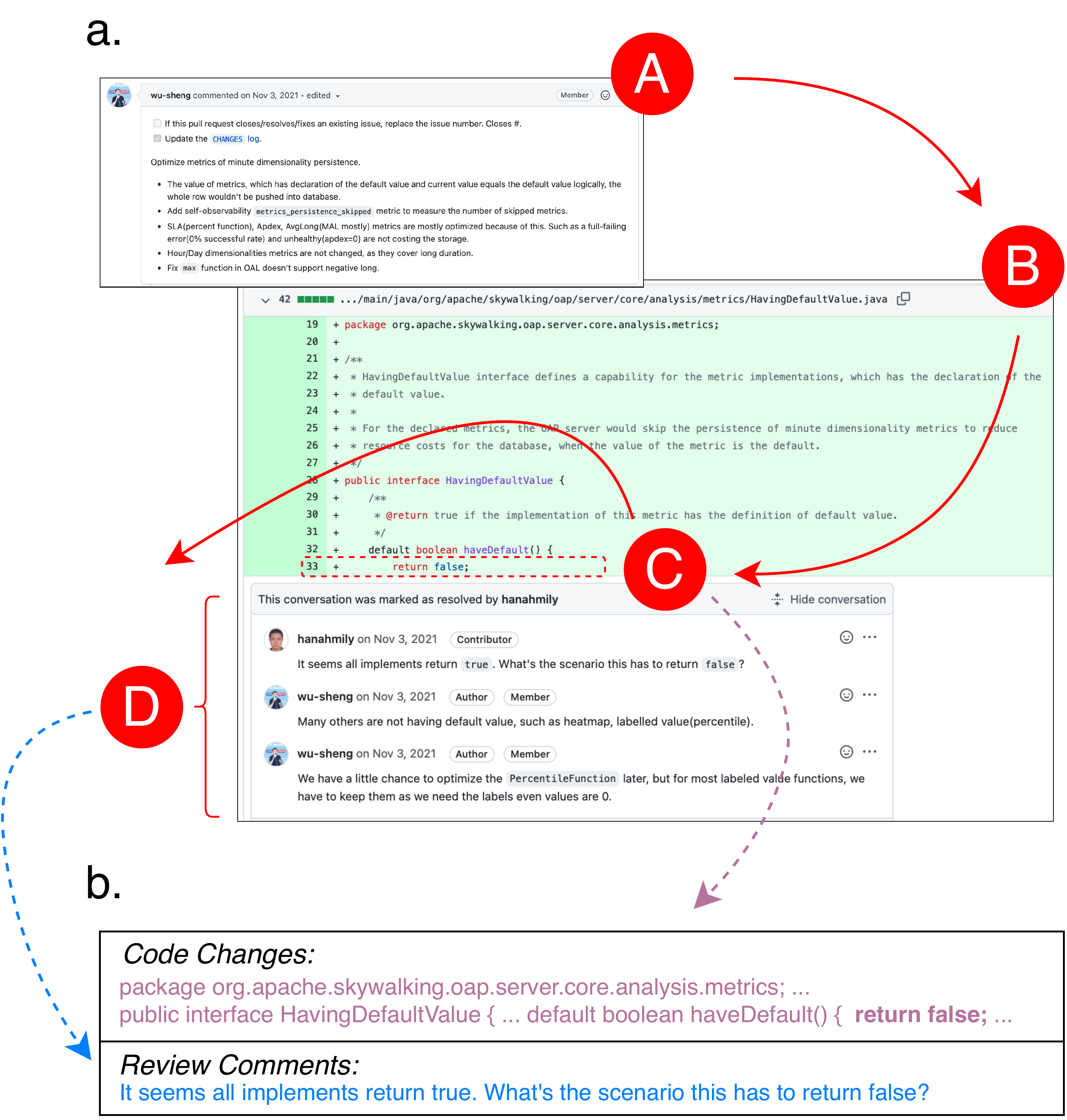}
  \caption{ An Example of Code Review on GitHub}
  \Description{a website view of code review on GitHub with index.}
  \label{fig:github}
\end{figure}
programming defects and improving the quality of software in early phase of development \cite{10.1145/2597073.2597076, 10.1145/3377811.3380385, 6606617}. Therefore, it is widely adopted nowadays in both open source and industry workflows \cite{6606642}.
 Reviewers need to thoroughly understand the source code during code review activities and leave review comments to interact with developers. Figure \ref{fig:github}.a shows an example of the review process. (A) First, developers create a pull request to submit code changes. (B) Then, reviewers receive the request and inspect the involved programming files. (C) Once they fully judge and disagree with some lines, (D) reviewers will leave messages and set up a conversation right behind them as in-line review comments. In summary, the assignment of reviewers is to double-check code changes, highlight review lines (bold font) and deliver useful review comments (Figure \ref{fig:github}.b).
 
 The nature of this cross-validation brings high quality to software \cite{10.1145/3474624.3477063}. For reviewers, they provide professional suggestions with their programming experience, which the author may not have been through \cite{7962371}—for example, a convenient coding style, a particular test case, or a complicated application scenario. For authors, they can learn from perspectives other than themselves and have a higher chance to perform efficiently \cite{7332454}. However, two factors limit the benefits:
 
 1) The usefulness of existing review comments remains uncertain. Czerwonka et al. found that only 15\% of review comments indicate a possible defect \cite{7202946}, and Bosu et al. reported that 34.50\% of review comments from five major projects of Microsoft are non-useful \cite{7180075}. Studies also proposed that low-quality review comments are considered useless and somehow mislead developers \cite{6606642, 7081827}. Even so, it seems still inevitable in practice. On the one hand, both authors' and reviewers' programming knowledge is still limited since they can't think over all possible conditions. On the other hand, it is difficult to assign review requests to correct reviewers due to that limitation \cite{tufano2021towards}. The assessment of useful review comments is controversial, too. Studies have argued that useful review comments can successfully identify functional issues or reflect other motivations, such as the social benefits of developers \cite{7180075, 6606617}. However, research by Rahman et al. reported that review comments could be defined as useful only if they trigger code changes within ten lines after reviewing \cite{7962371}. 
 
 2) The heavy involvement of human effort is annoying and time-consuming. Researchers argued that code review might be the lengthiest part of the development \cite{10.1145/2597073.2597076, 7202946}. Reviewers have to change work context unwillingly and fully understand a source code from others before commenting. For authors, they can't move forward until reviewers send their review comments back \cite{7202946}. Both of them are highly time-consuming in practice \cite{6606642}. On average, it costs developers 6 hours per week to handle code review issues \cite{bosu2013impact}. Furthermore, due to the low efficiency, more contributors are assigned to the review process. Around 20\% developers at Mozilla, a free software community, face heavy workloads, i.e., over ten patches/reviews per week beyond their own jobs \cite{7886977}. In Microsoft Bing, industrial projects can undergo approximately 3,000 code reviews per month \cite{10.1145/2491411.2491444}. In addition, developers have to spend an extra effort to work as reviewers every time they participate in an unfamiliar project.

Prior studies have made attempts to solve these two issues of review comments. Balachandran proposed \textit{Review-Bot}, which integrates the output of multiple static analysis tools to publish review comments automatically \cite{6606642}. Rahman et al. reported that their \textit{RevHelper} could directly predict whether review comments are useful or not \cite{7962371}. Though competitive, they work as an indirect integration and validation, not for the review itself. Tufano et al.'s latest study focused on automated code review with automatic code revision and implemented comments generation as a sub-task, too \cite{tufano2022using}. However, their method remains coarse-grained and only aimed at finding defects in code functions, not review lines.

In this paper, we explore an unlimited and efficient code reviewer and propose a novel approach, called \textbf{AUGER}, to generate review comments automatically with continuous training at the first step. We leverage pre-training models to synthesize effective manual review comments, thanks to its proven efficiency for big data in Natural Language Processing. From 11 influential open-source Java projects in GitHub\footnote{https://github.com/} (the largest source code host \cite{6976151}), we first collect those review pairs of code changes and review comments in 20K pull requests, which causes code revision. And then, we process data with heuristic methods and semantic augmentation in Data Preparation. With around 62K pieces of processed data, AUGER addresses the problem into three sub-processes: 1) Review Lines Tagging finds and highlights review lines that get revised after reviewing in code blocks with a unique leading review tag; 2) Cross Pre-training learns the inner correlations between code changes and review comments language with a masked language model from \textbf{T}ext-\textbf{T}o-\textbf{T}ext \textbf{T}ransfer \textbf{T}ransformer (T5); 3) Comments Generation fine-tunes the pre-trained model further with selected pairs and finally transfers code changes in the programming language to review comments in natural language. Our experiment results show that AUGER achieves 22.97\% in ROUGE-L and 4.32\% in Perfect Prediction, which outperforms baseline by 37.38\% and even 14 times. For efficiency, AUGER generates one piece of review comments as fast as 20 seconds. We also implement subsequent training and illustrate that AUGER can retrain freely in ablation. Moreover, we assess the usefulness of generated review comments with heuristic metrics from prior studies and present case study. We believe that AUGER has achieved our primary goal and is open to improvement in the future. The significant contributions of our paper are:

\begin{itemize}
\item We first formulate an issue of code review as a problem that generating review comments on manual-labelled problematic lines in code functions. To the best of our knowledge, this is the first study exploring an interactive automation by sharpening code changes.
\item To solve the problem, we propose AUGER, an synthetic and efficient generator to generate review comments automatically with instructive review tags, limitless training and immediate generation.
\item We evaluate AUGER on 1,088 test data from code review practice in GitHub, and it performs effectively in both automatic evaluation and heuristic assessment.

\end{itemize}

The rest of this paper is introduced as follows. Section 2 describes the background and definition of the problem. Section 3 illustrates the detailed approach. Section 4 presents the experiments. Section 5 shows the evaluation of AUGER. Section 6 introduces the related work. Section 7 illustrates threats to validity. Section 8 makes a conclusion.


The AUGER\footnote{https://gitlab.com/ai-for-se-public-data/auger-fse-2022} model and all experimental materials are publicly available to reproduce our results.

\begin{figure*}
  \centering
  \includegraphics[width=\linewidth]{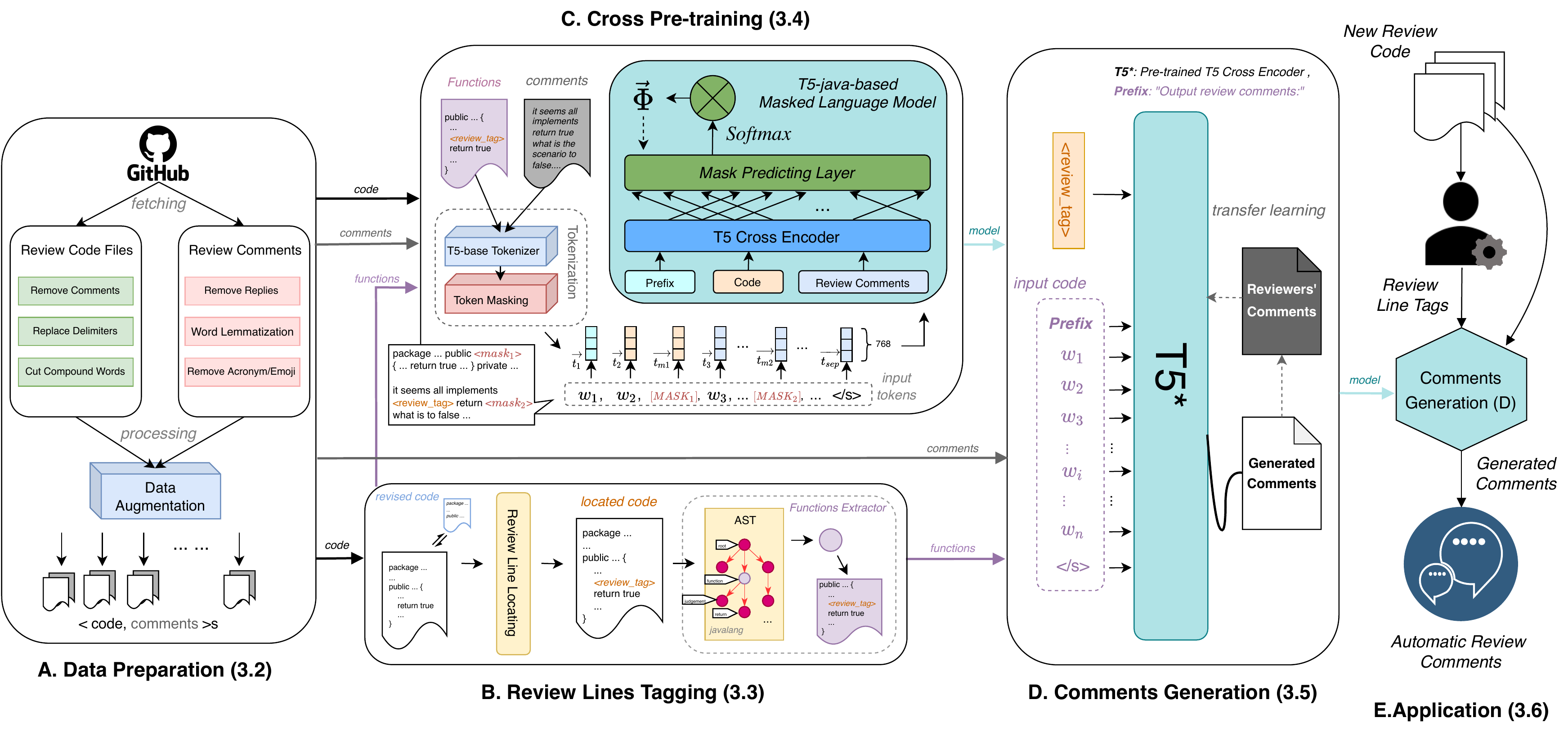}
  \caption{the Framework of AUGER}
  \Description{the framework of AUGER}
  \label{fig:auger}
\end{figure*}

\section{Background}

This section introduces the necessary background and the problem definition.

\subsection{Review Documents}

Modern code review is usually adopted in grand projects consisting of numerous documents \cite{10.1145/2597073.2597076, 6606642}. To simplify the illustration, we present involved review documents as several equations:

\begin{equation}
\begin{aligned}
  &D = \{d_1, d_2, ..., d_n\},\\
  &Sub\_{}Code = \{v_1^s, v_2^s, ..., v_{ns}^s\} \in D,\\
  &Rev\_Code = \{v_1^r, v_2^r, ..., v_{nr}^r\} \in D,
\end{aligned}
\label{equ:documents}
\end{equation}

The Equation \ref{equ:documents} represents the main materials employed during the course. As mentioned above, code review will not commence until a submission of code changes. Hence, given a universal $D$ of all documents $d_i$, we define these submitted code changes as $Sub\_{}Code$. After reviewing, versions of the code that submitters may revise is defined as $Rev\_Code$. $v^s$ and $v^r$ denotes code blocks divided by abstract syntax tree (AST) in these two texts, respectively.

\subsection{Review Process}

When review requests trigger, reviewers are assigned to inspect, analyze, and produce review comments in chronological order \cite{10.1145/2491411.2491444}. We describe the process with the following symbols:

\begin{equation}
\begin{aligned}
  &P = \{\vec{p_1}, \vec{p_2}, ..., \vec{p_m}\},\\
  &\vec{p_i} = <v_j^s, T, C, \vec{I}>
\end{aligned}
\label{equ:process}
\end{equation}

The series of processes is defined as a set $P$ in Equation \ref{equ:process}. Each element $p_i$ denotes one single reviewers' manipulation at that moment. For each $p_i$, the inspector notices a code block $v_j^s$ with possible issues in provided code changes, selects incorrect lines with a click, and leaves review comments $C$. To present it better, we abstract the click operation as a special leading tag $T$ to distinguish those highlighted lines in the text. Besides, the system records relevant messages $\vec{I}$ such as timestamp and developer id for each submission.

\subsection{Automatic Generation}

The automatic generation task is to automate the review process based on materials defined in the above sections. Thus, we formulate our key idea with two equations:

\begin{equation}
\begin{aligned}
  &\vec{R} = <Sub\_Code, P>,\\
  &P_{auto}:<v_j^s, T>\rightarrow C_{gen}
\end{aligned}
\label{equ:auto}
\end{equation}

In Equation \ref{equ:auto}, code review practice is first summarized as a combination of complete submitted code changes $Sub\_Code$ and reviewers' operation $P$ from Equation \ref{equ:documents} and Equation \ref{equ:process}. The automatic process $P_{auto}$ is defined as: a text generation function from code blocks $v_j^s$ and review line tags $T$ combination $<v_j^s, T>$, mapping to an output of generated review comments $C_{gen}$, like the manual one $P$. In other words, the motivation of AUGER is to act as an effective automatic substitution $P_{auto}$ for the human process $P$.

\section{Approach}

\subsection{Overview}
Figure \ref{fig:auger} presents the framework of AUGER, which has five abstract components: A. Data Preparation (3.2); B. Review Lines Tagging (3.3); C. Cross Pre-training (3.4); D. Comments Generation (3.5); E. Application (3.6).

Given a universe of documents $D = \{d_1, d_2, ..., d_n\}$ fetched from GitHub projects, Data Preparation selects and pre-processes the review code $Sub\_{}Code = \{v_1^s, v_2^s, ..., v_{ns}^s\}$ and review comments $C$. Review Lines Tagging highlights review lines with special tags $T$ where revision will commence, according to $Rev\_Code$. Cross Pre-training part learns the inner distribution of review text first, and then Comments Generation transfers the combination (code blocks $Sub\_{}Code$ with leading review tags $T$) into review comments $C$ automatically. After fully pre-training and fine-tuning, AUGER $P_{auto}$ can be applied to generate automatic review comments $C_{gen}$ from new code changes $Sub\_{}Code$ and manual tags $T$ in Application.

\begin{figure*}
  \centering
  \includegraphics[width=\linewidth]{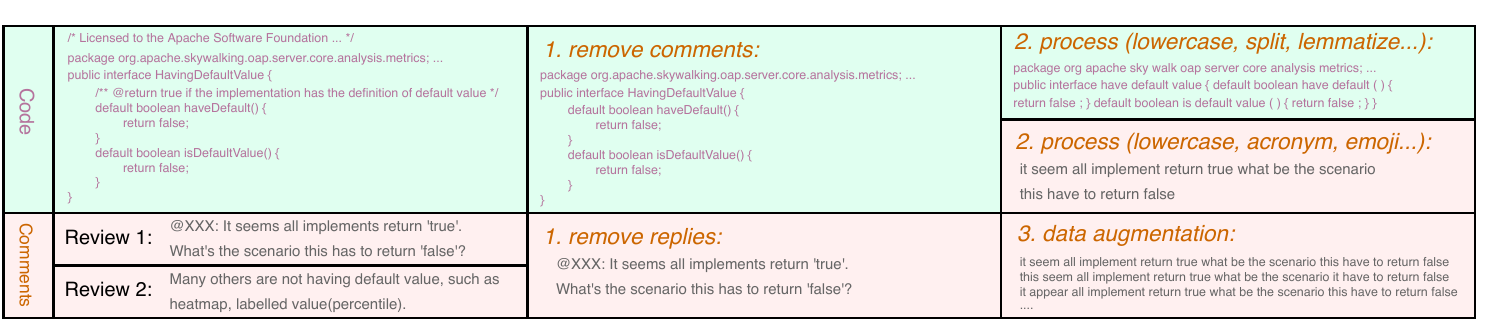}
  \caption{An Example of Data Preparation}
  \Description{an example of data preparation}
  \label{fig:datapreparation}
\end{figure*}

\subsection{Data Preparation}

1) Data Fetching: When fetching data from GitHub, we mainly follow the instruction of GitHub GraphQL API\footnote{https://docs.github.com/en/graphql/}. To ensure the effectiveness of review data with experienced review comments, we exactly fetch those review activities that indeed trigger code revisions \cite{7180075}. Besides, we remove those reviews not for Java code. After eliminating invalid and repetitive data, triplets \textit{<$Sub\_{}Code$, $Rev\_Code$, $C$>} are ready for further process. Figure \ref{fig:datapreparation} exhibits an example of Data Preparation: a pair of code changes and review comments input. For code text, we first remove source comments and then implement several processes such as splitting. Almost a same process is reimplemented for comments but an extra data augmentation afterwards.

2) Data Processing: We first check duplicates again to ensure that there is no in-train/in-test/cross-set duplicates causing data inflation \cite{allamanis2019adverse}. Then we remove nearly 85\%(67672) data consisting of two parts: 1) 40.77\%(32348) reviews where code revision or review comments include content beyond functions. In terms of function-related review activities, they are out-of-range and thus considered noisy. 2) 44.52\%(35324) “<sub\_code, comments>” pairs where code functions or review comments are shorter than three words or longer than 128. Some of them are considered noise, such as incomplete functions and useless replies like “fixed”. Besides, big functions are considered smelly \cite{beck1999bad} and too long for ML models, along with lengthy comments that include code blocks not in natural language. The rest 15\%(11672) are regarded as raw data and ready for next process. 

Then we convert all text into lowercase, split words and punctuation, and implement word lemmatization using NLTK\footnote{https://nltk.org/} toolkit ("seems" to "seem", "is" to "be" in Figure \ref{fig:datapreparation}). Then we process code and comments further so that models can quickly learn later.

For code text, we first totally remove source code comments with a heuristic filter. We preserve every punctuation except "." owing to its representation of structure knowledge. In Java, functions and variables are usually named within long and compound words. In this case, we split them into fine-grained sub-words with the help of WordNinja\footnote{https://github.com/keredson/wordninja/}. For example, from \textit{"HavingDefaultValue"} to \textit{"having default value"} in Figure \ref{fig:datapreparation}.

For comments, we erase all signs such as "@" in natural language. Then acronyms are replaced by their full names referring to Oxford Dictionary\footnote{https://public.oed.com/how-to-use-the-oed/abbreviations/} ("What's" to "What is" in Figure \ref{fig:datapreparation}). After that, We remove all punctuation to preserve semantic words. The process of code fragments in comments is as same as in code text. Last, we select the first 3 sentences when review comments are too long.

3) Data Augmentation: This process is widely applied to handle the lack of training data in NLP \cite{devlin2018bert, raffel2019exploring, joshi2020spanbert}. Though we have fetched enough raw data, the amount decreases sharply after being filtered. Besides, studies reported that data augmentation is able to expand the influence of core knowledge in the training \cite{devlin2018bert}. Hence, we boost triplets \textit{<$Sub\_{}Code$, $Rev\_Code$, $C$>} by 9 times with only increasing review comments $C$ using Wei et al. \cite{wei2019eda}.

\subsection{Review Lines Tagging}
Reviewers actually conduct reviewing on specific code lines. In practice, they highlight review lines first and then make review comments $C$ right after them. To be aligned with the fact, we synthesize this fine-grained review knowledge and reprocess the prepared code text to find out those review lines with a review tag $T$. Since we are only concerned about those code blocks indeed get revised afterward, this component is designed to remove irrelevant code blocks, too.

1) Review Lines Locating: In our implementation, we add a special token \textit{<review\_tag>} to the front of review lines. As shown in Equation \ref{equ:location}, after receiving code text from Data Preparation, we first compare original code changes $Sub\_{}Code = \{v_1^s, v_2^s, ..., v_{ns}^s\}$ and its revision $Rev\_{}Code = \{v_1^r, v_2^r, ..., v_{nr}^r\}$ to find out different lines $Diff$. Then we only select valid modifications $Diff_{valid}$ with a series of line changes. Subsequently, we set \textit{<review\_tag>} at the beginning of it to mark where revision happened. This process follows the practice that developers leave review comments on chosen lines for collaborators easy to understand. Although knowing where to comment is also a problem, it takes as fast as one inference when generating several comments in a batch. Hence, people can select all code lines they want and choose which results to learn or improve. What's more, if review tags are predicted automatically, an additional error is brought to the overall performance. 

\begin{equation}
\begin{aligned}
  &Diff = \{v^s | v^s \in Sub\_Code\}, \\
  &Diff_{valid} = \{v_i^s, v_{i+1}^s, ..., v_{i+k}^s\}, \quad k \in N
\end{aligned}
\label{equ:location}
\end{equation}

2) Functions Extractor: Studies have demonstrated that noise in input data can badly mislead the performance of language models \cite{khayrallah-koehn-2018-impact, dusek-etal-2019-semantic}. For safety, functions extractor is a tool to narrow the code range down to the least block $v_i^s$. First, we employ javalang\footnote{https://github.com/c2nes/javalang/} to fomulate code files into abstract syntax trees (AST). In AST, the structure of code is arranged in a hierarchy instead of a inclusion. For example, a root node is at the top, while other functions, judgments, and statements follow it. Similar to prior works \cite{tufano2021towards, mastropaolo2022using}, we extract code at the function height and select the one including target \textit{<review\_tag>}. 

\subsection{Cross Pre-training}
Recent studies confirm that the masked language approach in pre-training can facilitate the performance of models in many NLP tasks \cite{devlin2018bert, wang-etal-2020-pretrain}. Therefore, we synthesize review knowledge by pre-training a better contextual representation and bridging the relationship between code and comments with cross pre-training technique.

1) Tokenization: We pair processed comments $C$ (from Data Preparation) and functions $v_t^s$ (from Review Lines Tagging) into \textit{<$v_t^s$, $C$>}. As pre-training usually comes into effect with numerous inputs, we also integrate raw data from Data Preparation to increase \textit{<$v_t^s$, $C$>} directly. Then a T5-base tokenizer is employed to tokenize each of the text pairs (including \textit{<review\_tag>}) in one sentence. According to what BERT has proven effective in token masking \cite{devlin2018bert}, we randomly select 10\% words to mask.

2) Mask Language Model: In this part, we build a T5-java-based Masked Language Model to do language masking for pre-training. So far, there has been a lot of varieties of T5 models \cite{liu2021pre}. We choose a version of CodeTrans \cite{elnaggar2021codetrans} already trained on Java documents (from Hugging Face\footnote{https://huggingface.co/}). First, it works as a cross encoder by encoding both code and comment tokens in one sequence. For \textit{<review\_tag>}, we increase the embedding size by 1 to present it as a special token. Following the generation of T5, we write a prefix here as: "Generating review comments:" for all input data. Then we set a mask predicting layer to capture hidden states and output prediction vectors. Then, the softmax layer smooths it into probabilities $\vec{\Phi}$:

\begin{equation}
\begin{aligned}
  &\vec{\Phi} = <\phi_1, \phi_2, ..., \phi_{|V|}>, \quad 0 \leq \phi_1, \phi_2, ..., \phi_{|V|} \leq 1 \\
  & \phi_1 + \phi_2 + ... + \phi_{|V|} = 1
\end{aligned}
\label{equ:softmax}
\end{equation}

$\phi_1$, $\phi_2$, ..., $\phi_{|V|}$ respectively describes the possibility of the mask to be the first, the second, ..., the last word in the $|V|$ length vocabulary. Naturally, we choose the maximum as the prediction back to the model. The loss function in pre-training is defined as:

\begin{equation}
\begin{aligned}
  &Loss_{pre} = -\frac{1}{|M|} \sum_{i=1}^{|M|} \sum_{j=1}^{|V|} y_{ij}^m log(\phi_{ij}^m)
\end{aligned}
\label{equ:pretrain-loss}
\end{equation}

A \textit{Cross-entropy Loss} is applied to measure the difference between the prediction and the truth. Equation \ref{equ:pretrain-loss} accumulates products of each predicting probability $\phi_{ij}^m$ logarithm and golden label $y_{ij}^m$ (in vocabulary $V$ for all masked tokens $M$), and the purpose of Cross Pre-training process is to reduce the loss.

\subsection{Comments Generation}

In Comments Generation, we collect outputs from Data Preparation, Review Lines Tagging, and Cross Pre-training, and then transfer them into review comments products.

1) Text Encoder: Now we successfully have code blocks $v_t^s$ with review tags \textit{review\_tag} and a pre-trained model. Again, we use the tokenizer of T5 \cite{raffel2019exploring} to embed all text, including \textit{review\_tag} and the prefix. The T5 tokenizer encodes all text $w_i$ into 768-dimensional vectors and adds an special end token \textit{</s>}.

2) Transfer Learning: After fully pre-trained in Cross Pre-training, we further fine-tune the T5 cross encoder on text generation. Such generation is also known as transfer learning, where models trained on a dataset at a large scale are employed to solve a specific task \cite{raffel2019exploring, 9134370}:

\begin{equation}
\begin{aligned}
  &f^{D_{\delta_k}}(x_i) = \{P(y_i^D|x_i^D)|y_i^D \in Y^D, i = 1,2,...,|X^D|\}, \\
  &f^{T_{\tau_k}}(x_i) = \{P(y_i^T|x_i^T)|y_i^T \in Y^T, i = 1,2,...,|X^T|\}
\end{aligned}
\label{equ:trans}
\end{equation}

In Equation \ref{equ:trans}, $f^{D_{\delta_k}}$ denotes the predicted conditional distributions the model output for the task $\delta_k$ in domain $D$. $x_i^D$ is a domain observation in feature space $X^D$, while $y_i^D$ is defined as a specific truth label in $y_i^D$. The purpose of transfer learning is to utilize the knowledge in $f^{D_{\delta_k}}$ to improve $f^{T_{\tau_k}}$ effect on task $\tau_k$ in domain $T$. In this task, the ground truth here is the real comments from reviewers, which gets prepared in the first step. 

Next, the loss function of comments generation is defined as:

\begin{equation}
\begin{aligned}
  &Loss_{trans} = -\frac{1}{|S|} \sum_{i=1}^{|S|} \sum_{j=1}^{|V|} y_{ij}^s log(\phi_{ij}^s)
\end{aligned}
\label{equ:finetune-loss}
\end{equation}

Similar to the pre-training one, Equation \ref{equ:finetune-loss} accumulates the difference between each predicting probability $\phi_{ij}^s$ and golden label $y_{ij}^s$ for all generated sentences $S$. And for each sentence, the inference will finish at the prediction of the stop token or the max length.

\subsection{Application}

The Application is straightforward. We mainly leverage a fully trained model from Comments Generation as a synergy. When new code comes in modern code review, reviewers can freely select review lines first. After being tagged, this coded text is fed into the generation model and transferred into outputs. Finally, the generated comments are integrated into automatic review comments and returned to developers.

\section{Experimental Design}

\subsection{Research Questions}

According to our research purpose, the evaluation aims at answering the following research questions:

\begin{itemize}
\item \textbf{RQ1}: How effectively does AUGER perform to generate review comments automatically, after synthetic training?
\item \textbf{RQ2}: What is the contribution of sub-component designs to the overall performance in AUGER?
\item \textbf{RQ3}: To what extent is the usefulness of comments AUGER generated, compared with manual ones?
\end{itemize}

We will specifically discuss research questions in Section 5 with the evaluation of AUGER.

\begin{table}
  \caption{Data Source}
  \label{tab:source}
  \begin{tabular}{c|ccc}
    \toprule
    Repository &PRs(all) &PRs(suc) &Reviews(suc)\\
    \midrule
    Graal &4,200 &222 &904 \\
    Dubbo &9,500 &678 &1,447 \\
    Netty &12,000 &1,365 &4,938 \\
    Apollo &4,200 &98 &305 \\
    Flink &18,200 &4,759 &23,274 \\
    Kafka &11,605 &2,039 &10,544 \\
    Skywalking &8,350 &611 &1,985 \\
    Redisson &4,100 &49 &82 \\
    Bazel &14,500 &515 &1,651 \\
    Jenkins &6,100 &965 &3,379 \\
    ElasticSearch &82,200 &8,215 &30,735 \\
    \midrule
    Total &\textbf{174,955} &\textbf{19,516} &\textbf{79,344} \\
    \bottomrule
  \end{tabular}
\end{table}

\subsection{Data Source} 
To make sure the quality of the dataset, we select 11 notable Java repositories from GitHub in top 60 stars: \textit{Graal\footnote{https://github.com/oracle/graal/}} \textit{, Dubbo\footnote{https://github.com/apache/dubbo/}} \textit{, Netty\footnote{https://github.com/netty/netty/}} \textit{, Apollo\footnote{https://github.com/apolloconfig/apollo/}} \textit{, Flink\footnote{https://github.com/apache/flink/}} \textit{, Kafka\footnote{https://github.com/apache/kafka/}} \textit{, Skywalking\footnote{https://github.com/apache/skywalking/}} \textit{, Redisson\footnote{https://github.com/redisson/redisson/}} \textit{, Bazel\footnote{https://github.com/bazelbuild/bazel/}} \textit{, Jenkins\footnote{https://github.com/
jenkinsci/jenkins/}} \textit{, ElasticSearch\footnote{https://github.com/
elastic/elasticsearch/}}. What's more, these projects have at least 4,000 pull requests and 100 contributors. The detail of each repository is listed in Table \ref{tab:source}. PRs(all) denotes the number of pull requests successfully got from repositories, PRs(suc) denotes the number of pull requests that contain code review comments, and Reviews(suc) denotes valid review nodes in Java.

\subsection{Baselines}
To better answer RQ1, we compare the performance of AUGER with some state-of-the-art baselines. Since we formulate the task as a text generation, we choose three methods that have been proven capable to handle text generation and a latest study of code review \cite{boutaba2018comprehensive}:

\begin{itemize}
    \item LSTM (Long Short-Term Memory) is a classic type of neural network to store information over extended time intervals \cite{hochreiter1997long}. Its novel gate units lessen error backflow sharply and achieves a significant performance on natural language generation \cite{boutaba2018comprehensive}.
    \item COPYNET addresses text generation into "sequence to sequence" learning with copying mechanism, which replicates input segments selectively to outputs \cite{gu-etal-2016-incorporating}.
    \item CodeBERT is a pre-training model proposed by Microsoft, which can effectively solve many downstream tasks in programming language \cite{guo2020graphcodebert}.
    \item Recently, Tufano et al. made a study on code review automation and introduced a variant of T5 model to handle automatic code revision \cite{tufano2022using}. The model here we employed is their code-to-comment version.
\end{itemize}

\subsection{Metrics}
We use the ROUGE metric and Perfect Prediction rate to evaluate AUGER and its baselines. The ROUGE metric is widely used in Machine Translation evaluation \cite{lin-2004-rouge}. We employ ROUGE-1 and ROUGE-L score, which measure the overlap of words and the longest sequence between hypothesis and reference. Specifically, ROUGE-1 has precision, recall, and F1-score scores on word level. Precision measures the ratio of the correct number in prediction to the amount, recall measures the ratio of the correct number in inference to all samples in truth, and F1-score is a harmonic mean of Precision and Recall. Finally, Perfect Prediction is the rate of forecasts completely the same as ideal ones.

\begin{table}
  \caption{Training Hyperparameters}
  \label{tab:hyperparameter}
  \begin{tabular}{c|c|c}
    \toprule
     &Hyperparameter &Value\\
    \midrule
    \multirow{3}{55pt}{Pre-training Model}
        &Learning rate &1e-4 \\
        &Max input length &256 \\
        &Batch size &16 \\
    \midrule
    \multirow{6}{55pt}{Generation Model}
        &Learning rate &1e-3 \\
        &Length penalty &1 \\
        &Max input length &128 \\
        &Max output length &128 \\
        &Batch size &32 \\
    \bottomrule
  \end{tabular}
\end{table}


\subsection{Experiment Settings}
This section introduces the main parameters we set in all experiments.

1) Data Partition. After data processing, we preserve those review comments from 3 to 128 length. Same as code text. Then we split our dataset into 80\% training, 10\% validation and 10\% test, referring to Moreno-Torres et al. \cite{6226477}.

2) Pre-training Model. We employed a base T5 model fine-tuned on Java Code Documentation Generation Multitasks \cite{elnaggar2021codetrans}, and the model was trained in 60,000 steps and evaluated at every 6,000 step in pre-training. Some key hyperparameters are listed in Table \ref{tab:hyperparameter}.

3) Generation Model. Likewise, we selected the same T5 model to do the generation, and the model was trained in 28,000 steps and evaluated at every 2,000 step in fine-tuning. Table \ref{tab:hyperparameter} presents some key hyperparameters we set.

4) Experiment Environment. We implemented all training with NVIDIA GeForce RTX 3090 GPU and CUDA 11.4\footnote{https://developer.nvidia.com/cuda-toolkit/}, and it cost 12 hours and 5 hours to pre-train and fine-tunes AUGER, respectively.

5) Implementation of baselines. We leveraged LSTM and CopyNet built by AllenNLP\footnote{https://allenai.org/allennlp}. The CodeBERT model comes from Hugging Face\footnote{https://huggingface.co/} and the code-to-comment T5 variant comes from the repository of Tufano et al. \cite{tufano2022using}.

\section{Evaluation}

\subsection{RQ1: Performance of generating comments}

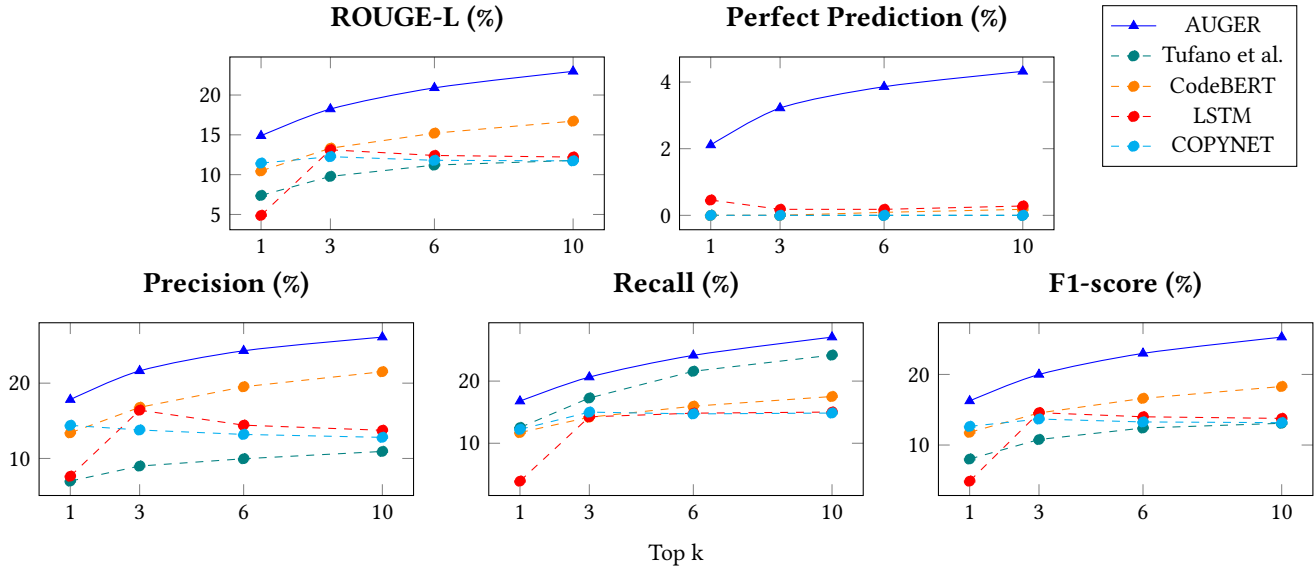
\begin{figure*}
\begin{tikzpicture}
	\begin{groupplot}[
	group style={group size=4 by 1}, 
	height=4.4cm, 
	width=\textwidth,legend columns=1,
	legend style={at={(2.59,1.3)},anchor=north},
	title style = {font=\Large},
	xlabel style = {font=\normalfont}
	]
		\nextgroupplot[group/empty plot, width=3.8cm]
		\nextgroupplot[title=\textbf{ROUGE-L (\%)}, height=3.88cm, width=6.56cm, xtick={1,3,6,10}]
		    \addplot[smooth,mark=triangle*,blue] coordinates {(1,14.88) (3,18.24) (6,20.90) (10,22.97)};
		    \addlegendentry{AUGER}
		    \addplot[dashed,mark=*,teal] coordinates {(1,7.39) (3,9.78) (6,11.20) (10,11.79)};
		    \addlegendentry{Tufano et al.}
		    \addplot[dashed,mark=*,orange] coordinates {(1,10.47) (3,13.32) (6,15.21) (10,16.72)};
		    \addlegendentry{CodeBERT}
		    \addplot[dashed,mark=*,red] coordinates {(1,4.89) (3,13.12) (6,12.41) (10,12.21)};
		    \addlegendentry{LSTM}
		    \addplot[dashed,,mark=*,cyan] coordinates {(1,11.45) (3,12.26) (6,11.80) (10,11.73)};
		    \addlegendentry{COPYNET}
		\nextgroupplot[title=\textbf{Perfect Prediction (\%)}, height=3.88cm, width=6.56cm, xtick={1,3,6,10}]
		    \addplot[smooth,mark=triangle*,blue] coordinates {(1,2.11) (3,3.22) (6,3.86) (10,4.32)};
		    \addplot[dashed,mark=*,teal] coordinates {(1,0.0) (3,0.0) (6,0.0) (10,0.0)};
		    \addplot[dashed,mark=*,orange] coordinates {(1,0.0) (3,0.0) (6,0.09) (10,0.18)};
		    \addplot[dashed,mark=*,red] coordinates {(1,0.46) (3,0.18) (6,0.18) (10,0.28)};
		    \addplot[dashed,,mark=*,cyan] coordinates {(1,0.0) (3,0.0) (6,0.0) (10,0.0)};
		\nextgroupplot[group/empty plot]
	\end{groupplot}
\end{tikzpicture}
\begin{tikzpicture}
	\begin{groupplot}[
	group style={group size=3 by 1}, 
	height=5cm, 
	width=\textwidth, 
	title style = {font=\Large},
	xlabel style = {font=\normalfont}
	]
		\nextgroupplot[title=\textbf{Precision (\%)}, height=3.88cm, width=6.56cm, xtick={1,3,6,10}]
		    \addplot[smooth,mark=triangle*,blue] coordinates {(1,17.81) (3,21.63) (6,24.30) (10,26.11)};
		    \addplot[dashed,mark=*,teal] coordinates {(1,6.99) (3,8.99) (6,9.96) (10,10.93)};
		    \addplot[dashed,mark=*,orange] coordinates {(1,13.40) (3,16.79) (6,19.52) (10,21.52)};
		    \addplot[dashed,mark=*,red] coordinates {(1,7.64) (3,16.41) (6,14.42) (10,13.74)};
		    \addplot[dashed,,mark=*,cyan] coordinates {(1,14.40) (3,13.78) (6,13.19) (10,12.80)};
		\nextgroupplot[title=\textbf{Recall (\%)}, xlabel=Top k, height=3.88cm, width=6.56cm, xtick={1,3,6,10}]
		    \addplot[smooth,mark=triangle*,blue] coordinates {(1,16.77) (3,20.64) (6,24.14) (10,27.07)};
		    \addplot[dashed,mark=*,teal] coordinates {(1,12.53) (3,17.28) (6,21.58) (10,24.18)};
		    \addplot[dashed,mark=*,orange] coordinates {(1,11.72) (3,14.19) (6,15.96) (10,17.52)};
		    \addplot[dashed,mark=*,red] coordinates {(1,3.91) (3,14.25) (6,14.84) (10,15.02)};
		    \addplot[dashed,,mark=*,cyan] coordinates {(1,12.23) (3,15.01) (6,14.72) (10,14.86)};
		\nextgroupplot[title=\textbf{F1-score (\%)}, height=3.88cm, width=6.56cm, xtick={1,3,6,10}]
		    \addplot[smooth,mark=triangle*,blue] coordinates {(1,16.24) (3,19.99) (6,22.98) (10,25.28)};
		    \addplot[dashed,mark=*,teal] coordinates {(1,8.02) (3,10.78) (6,12.41) (10,13.08)};
		    \addplot[dashed,mark=*,orange] coordinates {(1,11.80) (3,14.54) (6,16.61) (10,18.29)};
		    \addplot[dashed,mark=*,red] coordinates {(1,4.89) (3,14.58) (6,14.00) (10,13.77)};
		    \addplot[dashed,,mark=*,cyan] coordinates {(1,12.64) (3,13.68) (6,13.27) (10,13.17)};
	\end{groupplot}
\end{tikzpicture}
\caption{the Experiment Results on Top k Generation.}
\label{fig:experiment}
\end{figure*}

\begin{figure}
  \centering
  \includegraphics[width=0.9\linewidth]{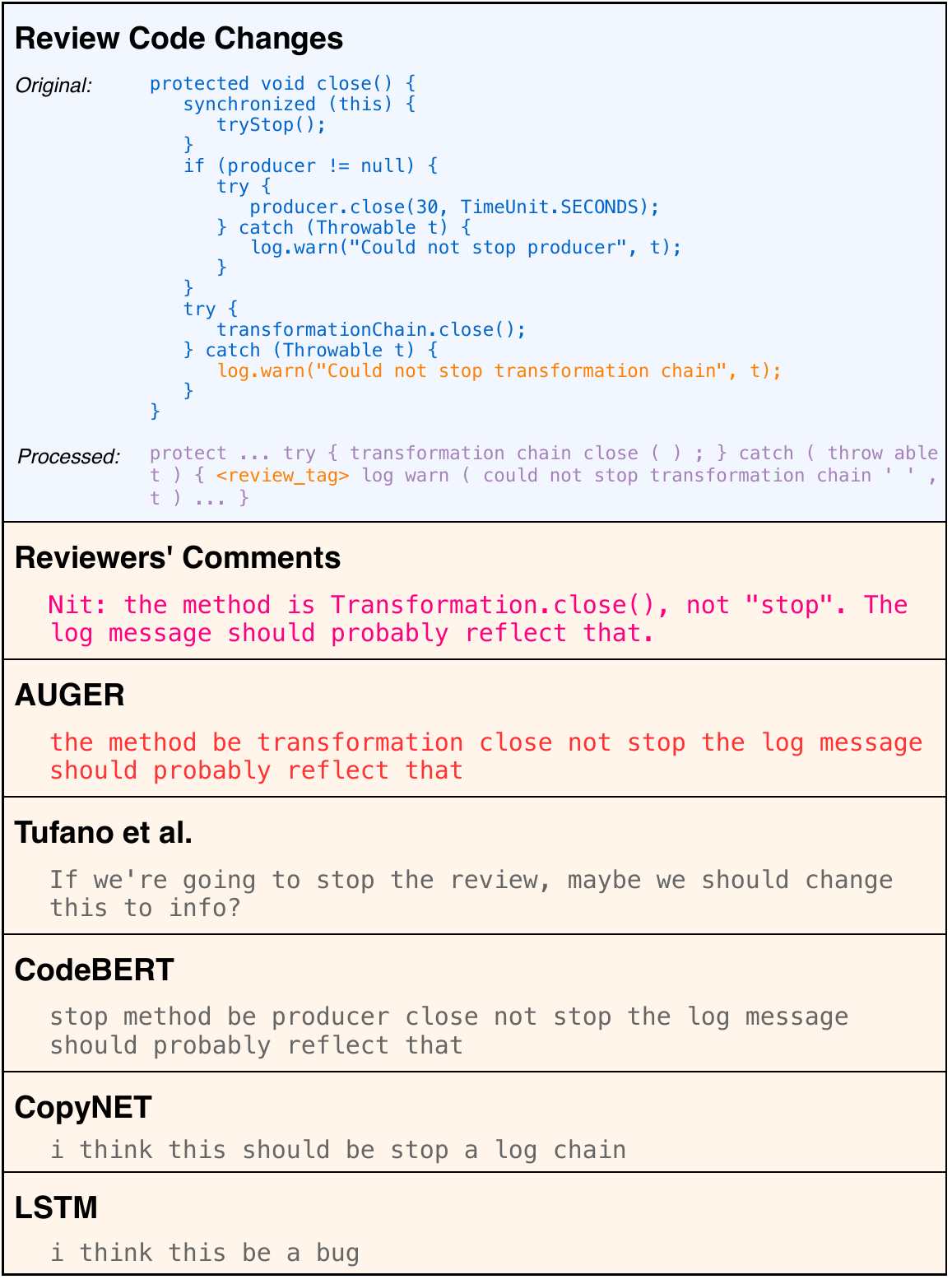}
  \caption{an Example of AUGER}
  \Description{an example of code review}
  \label{fig:test}
\end{figure}



The results in Figure \ref{fig:experiment} show that AUGER outperforms baselines in every situation. We mainly focus on the Top-10 scores on every metric since they are theoretically the best for 1 to 10. In Top-10, AUGER achieves 22.97\%, 4.32\%, 26.11\%, 27.07\%, 25.28\% on ROUGE-L, Perfect Prediction, Precision, Recall, and F1-score, and outperforms the best baseline by 37.38\%, 14 times, 21.33\%, 54.51\%, and 38.22\%, respectively. It indicates that our approach can generate review comments more precisely than baseline when fully trained, whether in a long statement or word similarity. In other cases shown in Figure \ref{fig:experiment}, AUGER also achieves apparent outperformance from 11.95\% up to tens of times as well.

We thoroughly analyze the results and explain the advantage of AUGER from two perspectives: 1) AUGER performs effectively on knowledge synthesis and text generation thanks to the advanced pre-training and transfer learning techniques. They provide a chance to learn knowledge from data at a large scale, whatever the language is. In that case, T5 is one of the most outstanding models. It outperforms traditional deep learning models like LSTM, CopyNET, BERT on multi-tasks besides code-to-comment generation. 2) Especially, we address review comments generation into a fine-grained task compared with prior works, which inputs highlighted lines led by a review tag within code blocks, not functions themselves. Thus, Review Lines Tagging helps AUGER make pertinent review comments related to key code lines. The benefit will be proven in ablation (Section 5.2), too.

In addition, it is worth noting that the Perfect Prediction rate of all methods is low. There are two likely causes: 1) Perfect Prediction is indeed a strict metric that requires complete equality between inference and ground truth; 2) methods have limitations on solving this task. LSTM and CopyNET directly implement training on both programming and natural language, which, according to their output, mainly captures the semantics in comments. CodeBERT and Tufano et al. pre-train themselves on various data, including other programming or code-to-code language, which brings uncontrollable noise to their performance on this task. What's more, Tufano et al. trained their T5 without review tags, and it may mislead the model to irrelevant lines. Still, it struggles due to their defect-oriented dataset construction, too. LSTM and CopyNET perform worse for that k bigger than 3. It is mainly caused by the overfitting they suffer from when handling text in different languages.

For efficiency, AUGER automatically generates one piece of review comments in 20 seconds on average. Compared with manual ones made in minutes and hours, AUGER highly alleviates the burden of review process, i.e., inspecting, analyzing and commenting. What's more, AUGER works as a non-resting reviewer with immediate feedback, from which developers can receive review comments just in time. It effectively saves both developers and reviewers time of waiting due to the discrepancy of assignment.

We also present an example in Figure \ref{fig:test} for intuitive comparison. An instance of review code changes to input and review comments from all baselines are listed in Figure \ref{fig:test}. The original code changes declare a protected function "close()" and the reviewer highlights a code line that monitors a statement "transformationChain.close()" and logs a warning message whenever the statement fails. Data Preparation and Review Lines Tagging will process the code into a code sentence as input. The second line shows that human reviewers find defects that the operation in the statement is "Transformation.close()" instead of "Transformation.stop" and make comments to point it out. Taking it as a reference, we compare each auto-output of machine techniques. AUGER performs the best and successfully find the defect by generating comments as same as a human. Comments from Tufano et al. and CodeBERT are in review language but confused. Although CopyNET and LSTM capture the defect on review lines, the outputs are ambiguous and useless opinions for submitters.

\begin{tcolorbox}
\textbf{Answering RQ1:} AUGER is capable to synthesize review knowledge and outperforms all baselines on five metrics when generating review comments automatically.
\end{tcolorbox}

\subsection{RQ2: Ablation experiment}

\begin{table}
  \caption{the Ablation Results}
  \label{tab:ablation}
  \begin{tabular}{c|ccccc}
    \toprule
    Methods &ROUGE-L &Perfect Prediction\\
    \midrule
    T5 base &22.01\% &3.95\% \\
    T5 java &22.41\% &4.14\% \\
    \midrule
    AUGER -<review\_tag> &21.47\% &3.31\% \\
    AUGER -pretraining &22.91\% &3.95\% \\
    AUGER* &\textbf{23.93\%} &4.04\% \\
    AUGER &22.97\% &\textbf{4.32\%} \\
    \bottomrule
  \end{tabular}
\end{table}

In the ablation experiment, we first compare AUGER with two fundamental T5 versions widely used in research. These two models are from Hugging Face\footnote{https://huggingface.co/}. We trained the common T5-base and the Java pre-trained T5-base model on our dataset and listed their results on the upper half of Table \ref{tab:ablation}. These two models are the initial version that are fine-tuned with review tags and processes like word cut, rather than the one Tufano et al. trained with their raw data. The outperformance proved that AUGER could solve the problem more thoroughly than a simple fine-tuning on T5 models (-4.18\% and -2.44\%). We further remove Review Lines Tagging and Cross Pre-training parts, and the results are shown on the lower half (AUGER -<review\_tag>, AUGER -pretraining). Similarly, both removals cause a decline in AUGER's performance (-6.53\% and -0.26\%). 

To illustrate the unlimited synergy better, we warm boost AUGER at the half of training data and implement subsequent training for the rest. The result in the test (AUGER* in Table \ref{tab:ablation}) shows nearly no efficiency loss and even an increase of ROUGE-L compared with the original one (AUGER). We believe the reason is that all data is randomized and more text related to the test occurs earlier in training steps. However, the overall performance remains stable and indicates that AUGER has the ability of training its synergy further.

The results illustrate that each component is contributive to the overall performance. The reasons are as follows: 1) As mentioned in Section 5.1, Review Lines Tagging narrows the scope down to highlighted review lines instead of a complete code file or function and saves models from deviated comprehension. 2) Cross Pre-training thoroughly trains the model with multiple steps and guides it to build the relationship between two different modalities, i.e., programming language and natural language. The improvement of these two modules is essential for AUGER. Notably, the warm boosting of AUGER (AUGER*) performs similarly, even better than AUGER. It indicates that AUGER is capable of working as an unlimited reviewer with subsequent training steps. As mentioned in Section 1, the scalability breaks the limitation of practical code review and sharply reduces the manual effort to grasp unfamiliar programming knowledge.


\begin{tcolorbox}
\textbf{Answering RQ2:} In the framework of AUGER, every component counts for the overall achievement and supports AUGER to train further.
\end{tcolorbox}

\subsection{RQ3: The Assessment of Usefulness}

\begin{figure}
\begin{tikzpicture}
\begin{axis}[
        xbar,
        xmin=0,
        y=0.45cm,
        width = 6.5cm,
        height = 1.8cm,
        bar width=8pt,
        xlabel={Percentage (\%)},
        nodes near coords,
        symbolic y coords={Track Rationale, Avoid Build Breaks, Team Assessment, Share Code Ownership, Improving Dev Process, Knowledge Transfer, Team Awareness, Alternative Solutions, Code Improvement, Finding Defects},
        ytick = data,
        enlarge x limits={value=0.2,upper},
        legend pos=south east,
        legend image post style={scale=2.1},
        legend style={row sep=8pt, column sep=8pt}
    ]
    \addplot[fill=black] coordinates { 
    (2,Track Rationale) (2,Avoid Build Breaks) (6,Team Assessment) (6,Share Code Ownership) (6,Improving Dev Process) (12,Knowledge Transfer) (22,Team Awareness) (34,Alternative Solutions) (46,Code Improvement) (60,Finding Defects) };
    \end{axis}
\end{tikzpicture}
\caption{Motivation Reflection of Generated Comments}
\label{fig:typenum}
\end{figure}
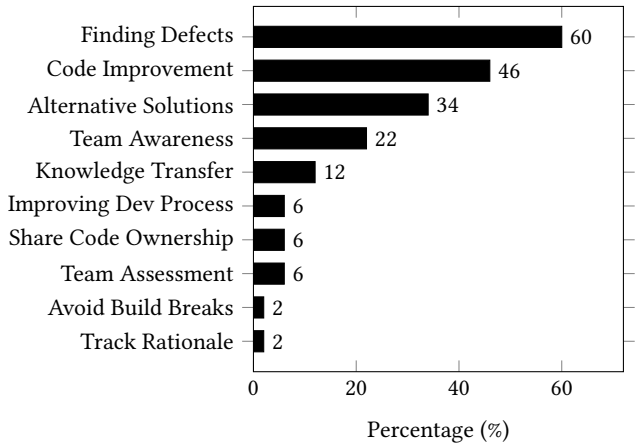

\begin{figure}
  \centering
  \includegraphics[width=\linewidth]{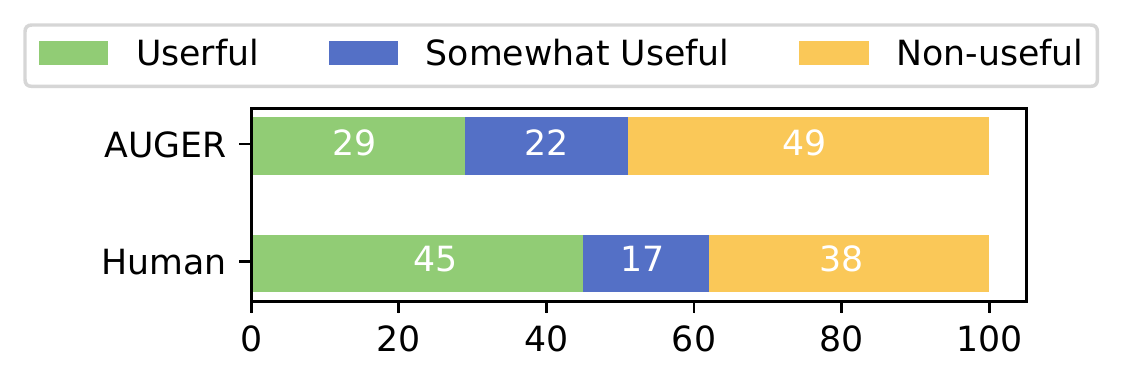}
  \caption{the Percentage of Useful Review Comments}
  \Description{the usefulness}
  \label{fig:casestudy}
\end{figure}

\begin{figure*}
  \centering
  \includegraphics[width=\linewidth]{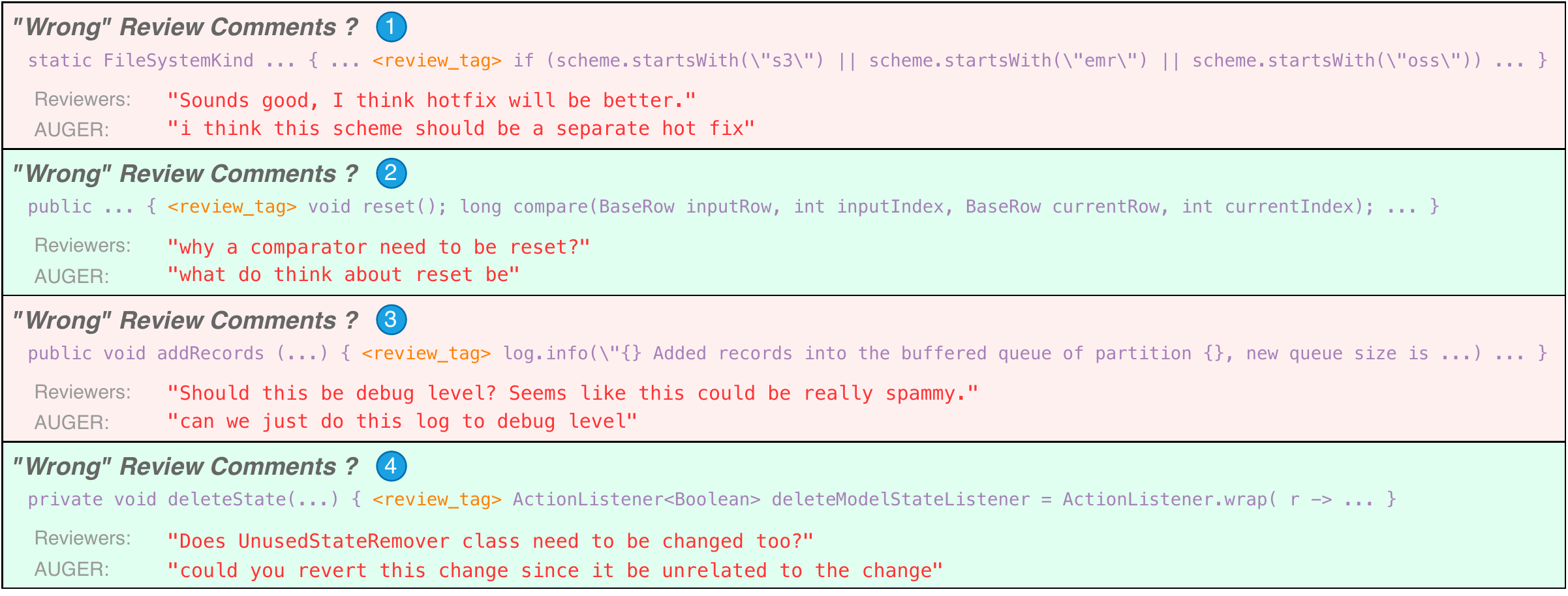}
  \caption{Examples in Case Study}
  \Description{comments types}
  \label{fig:types}
\end{figure*}

Subsections 5.1 and 5.2 demonstrate the ability of AUGER to handle the issues of code review practice being restricted and time-consuming.
Furthermore, to ensure the effectiveness of AUGER, we sample 100 (1\%) of the best (not perfect) inference in the test and evaluate the usefulness. As mentioned in Section 1, AUGER is our first step and aimed at commenting as effectively as human reviewers. Hence, according to Rahman et al. \cite{7962371}, we mainly compare it with manual review comments considered useful, which, in other words, indeed trigger source code changes on the author's side after reviewing (described in Section 3.2). The consideration follows the definition of Rahman et al \cite{7962371}. In that case, the effectiveness has been evaluated early in the last two subsections (5.1 and 5.2) on automatic metrics that calculate the similarity between the generation and ground truth.

However, auto-metrics only calculate word similarity between products and ground truth, but no semantics. However, review comments should be understandable. We found that most products of traditional models, i.e., LSTM and CopyNet, are non-readable or suffer from repetition. They are far from use even though they are competitive in auto-metrics. AUGER aims to generate readable comments and make sense for code review, just like humans. Hence, though much difficult, we still compare the usefulness with manual ones, the only qualified reference.

We first employ the standard proposed by Baccheli et al. \cite{6606617}. It reported that review comments are supposed to reflect ten of developers' motivations to drive code review. Some of them are for programming issues, while other motivations are for social benefits in several aspects. In that case, we analyze how effectively AUGER performs usefully to reflect them and present the result in Figure \ref{fig:typenum}. The 100 samples successfully cover every aspect of valuable review comments as expected. The distribution is similar to Baccheli et al. of 570 actual data, too \cite{6606617}: Finding Defects and Code Improvement are the most common reasons (60\% and 46\%) why developers drive code review, but other factors matter as well, such as Team Awareness of social benefits.

Next, we implement a classification of samples into three types: \textit{useful, somewhat useful, non-useful}. It refers to the empirical study of Bosu et al. \cite{7180075}, in which they list several assessments for each type. Compared with manual ones, we also classify the ground truth of these 100 samples in the same criterion. For AUGER, it generates comments 29\% useful, 22\% somewhat useful and 49\% non-useful. For reviewers, they perform better: 45\% useful, 17\% somewhat useful, and 38\% non-useful. It seems reviewers present more accurate review comments when inspecting code changes themselves and outperform AUGER by 55.17\% for useful comments. However, it decreases to 28.95\% when calculating the overprediction of non-useful review comments. Especially, the criterion was built on the analysis of manual hand-writings and overweight the diversity of comment language. For example, they assumed that review comments were somewhat useful whenever they started with a "nitpick". But, AUGER characterizes the word as a special feature in review comments text and generate it frequently. What's more, Bosu et al. judged those of social benefits non-useful such as the impressive building of developers. Still, Baccheli et al. argued that it is of great importance for team collaboration \cite{6606617}.


\subsection{Case Study}

In this section, we conduct a case study to understand the effectiveness of AUGER to generate review comments better. We inspect the 100 samples in Section 5.3 and further analyze those not as perfect as humans when calculating Perfect Prediction. Among 96 "wrong" review comments, we find 3 "nearly perfect" with only a different word. There are 4 "wrong" review comments delivering the same message as the manual one. We report them in Figure \ref{fig:types}.

In Case \textbf{No.1}, reviewers judge the code change of "if" judgment practical and advise to raise a hotfix here instead of the merge request. Though not perfect, AUGER also reports a hotfix and successfully captures the critical variable "schema".

In Case \textbf{No.2}, reviewers propose a question about the meaning of the "reset" statement. Similarly, AUGER asks the author why to do "reset" here in another accent.

In Case \textbf{No.3}, reviewers point out that the "log" statement should be in a "debug level" since it's really "spammy". For AUGER, though not explaining the reason, it also reports the main issue and suggests setting the "log" to "debug level".

In Case \textbf{No.4}, reviewers think that the change of "UnusedStateRemover" is not necessary here. AUGER fails to point it out but still suggests removing the change on review lines similarly.

In conclusion, we totally scan the 100 review comments AUGER generates in the test and find 3 "nearly perfect" and 4 "perfect in meaning". In that case, we can estimate a latent improvement of AUGER's performance when fully analyzed.

\begin{tcolorbox}
\textbf{Answering RQ3:} Referring to two criteria of prior works, review comments generated by AUGER are nearly as useful as manual ones. It also reveals an underestimate for those "wrong" samples.
\end{tcolorbox}


\section{Related Work}
We focus our discussion on (i) studies on modern code review, (ii) pre-training models, and (iii) approaches to generating source code comments. 


\textbf{Modern Code Review}. 
Several studies tried to automate modern code review process due to its time-consuming nature \cite{shull2008inspecting, kononenko2016code}. Balachandran's method delivers comments automatically with the integration of multiple static analysis tools \cite{6606642}. Gupta et al. proposed DeepCodeReviewer (DCR), which uses deep learning to learn review knowledge and predict an ideal one from a repository \cite{gupta2018intelligent}. Shi et al. tried to predict the approval of the revised code and proposed DACE with the performance of 48\% F1-score \cite{shi2019automatic}. Moreover, Tufano et al. recommended code changes directly using their automatic models with the 30\% perfect prediction on their dataset \cite{tufano2021towards}. 

AUGER, compared to the techniques discussed above, is able to generate review comments with given code changes and provide human-like support to modern code review.

\textbf{Pre-training Models}. Pre-training models have achieved remarkable success in natural language processing at present \cite{blitzer2006domain, sarzynska2021detecting, radford2018improving, howard2018universal}. Devlin et al. introduced a simple but powerful model named BERT (Bidirectional Encoder Representations from Transformers) to outperform others on eleven NLP tasks with large margin. The model even surpassed human's performance in challenging areas \cite{devlin2018bert}. Following it, models trained in two stages have sprang up, i.e., learning representations at pre-training and fine-tuning on downstream tasks, and they are open to subsequent training. For example, RoBERTa \cite{liu2019roberta}, XLNET \cite{yang2019xlnet}, and GPT \cite{brown2020language}. In addition, models pre-trained on software texts become popular, with the potential to solve tasks like bug reporting \cite{tabassum-etal-2020-code, ciborowska2021fast, shi2021ispy}.

Owing to the effectiveness of this pattern, recent studies work on exploring a framework for diverse unsupervised text data. It is also known as transfer learning in terms of implementation \cite{radford2019language, shazeer2018mesh, huang2019gpipe, keskar2019ctrl}. Raffel et al. formulated all text-based language problem as a text-to-text problem and proposed a T5 model at large scale \cite{raffel2019exploring}. The model is designed based on the transformer architecture open to complicated input with self-attention layers, instead of RNNs or CNNs \cite{vaswani2017attention, cheng2016long}. By combining the knowledge from exploration with scale and new corpus, T5 characterized many tasks into text-to-text and achieved state-of-the-art results on benchmarks, including summarization, question answering and text classification \cite{raffel2019exploring}.

\textbf{Source Code Comments Generation}. Studies on code comments and comprehension of programs can be traced back to 1980s \cite{10.1145/382208.382523, 10.1145/1858996.1859006, 10.5555/800078.802534}. During software maintenance, good comments are of great importance to program understanding since developers are able to learn the meaning of code at ease with the help of this descriptive language \cite{tenny1988program}. However, developers sometimes do not comment their code adequately due to extra efforts, lack of relevant knowledge or overlooking the importance of code comments \cite{8778714}. Study reported that developers have to spend up to 59\% time on these manual activities, which limits the efficiency of software development heavily \cite{xia2017measuring}.

Automatic generation of source code comments just started in the last decade. The early methods focused on information retrieval, such as VSM algorithms \cite{6062165, 5645482}, clone detection \cite{7081848}, and LDA algorithms \cite{movshovitz2013natural}. After that, with the prosperity of deep learning, Hu et al. \cite{8973050} and Wan et al. \cite{10.1145/3238147.3238206} proposed their automatic models with deep neural networks. Recently, pre-training models including CodeBERT \cite{feng-etal-2020-codebert} and T5 \cite{elnaggar2021codetrans} have been proved to outperform the state-of-the-art on comments generation.

Distinct from code comments that describe source code function, review comments are professional knowledge to reflect inappropriate code lines. Hence they have different focuses. Besides, all information code comments have to capture is just from its source code text. However, it requires numerous practices to make useful review comments. We implement similar code-to-comments generation but different in terms of programming comprehension.

\section{Threats to Validity}

\textbf{Construct validity:} 
We use the original review comments written by reviewers as ground truth, assuming that they cause code revision. However, these comments have no guarantee to make sense for code improvement. For example, developers change their code by themselves right after code review activities. Therefore, datasets may contain some suboptimal review comments and affect the evaluation because we measure Perfect Prediction rate that only considers words entirely equal to manual ones. To partially address the threat, we manually analyzed a sample of non-perfect predictions in case study, calculating the percentage of valuable ones from different references.

\textbf{Internal validity:} A study shows that the impact of hyperparameters on T5 models remains unknown \cite{raffel2019exploring}. In that case, we fully explore prior works and follow them to set hyperparameters in our approach. We acknowledge that a few of other settings may cause better performance, which is also a part of our future study.

\textbf{External validity:} Although the dataset features thousands of instances, we limited our experiments to Java projects. Hence, we do not claim the generalizability for other programming languages. However, we select the notable systems with high-quality code reviews to train our model. Future work will further explore whether AUGER can learn knowledge of projects in different programming languages.

\section{Conclusion}

Since code review practice suffers from the limitation of individual collaboration, this paper proposed a model named AUGER, to generate review comments with pre-training T5 models automatically. We first fetch 79,344 Java reviews in Github and heuristically clean 85.29\% noisy data considered useless or irrelevant, such as regular replies. Then we build a framework leveraging Text-to-Text Transfer Transformers (T5) models to automatically integrate and generate review comments. The synergy captures the relationship between code and review language effectively and shows better performance than baselines and high efficiency to immediate feedback. The model is capable to train further and cover unfamiliar programs more freely than individuals. Several criteria from prior studies are also employed to assess the generation as useful human-like review comments to some extent. 

We will further explore a language-agnostic implementation and a broad application to collaborate with professionals reviewing new systems in future work.

\section{Acknowledgments}

We sincerely appreciate the valuable feedback from the anonymous reviewers. This work was supported by the Strategy Priority Research Program of Chinese Academy of Sciences (No.XDA20080-
200), the National Key R\&D Program of China (No.2021YFC3340204) and Chinese Academy of Sciences-Dongguan Science and Technology Service Network Plan (No. 202016002000032).

\bibliographystyle{ACM-Reference-Format}
\bibliography{templet}

\end{document}